\documentclass{iau}
\usepackage{graphicx}

\title[Interchange of alternating amplitudes] 
{On the interchange of alternating-amplitude pulsation cycles}

\author[E. Plachy, L. Moln\'ar, Z. Koll\'ath, J. M. Benk\H{o}, K. Kolenberg]   
{E. Plachy$^1$, L. Moln\'ar$^{1,2}$, Z. Koll\'ath$^{2,1}$, J. M. Benk\H{o}$^1$ \\ \and K. Kolenberg$^3$}

\affiliation{$^1$Konkoly Observatory, Konkoly Thege Mikl\'os \'ut 15-17, H-1121, Budapest, Hungary \\ email: {\tt eplachy@konkoly.hu} \\[\affilskip]
$^2$  Institute of Mathematics and Physics, Savaria Campus, University of West Hungary \\ H-9700 Szombathely, K\'arolyi G\'asp\'ar t\'er 4, Hungary \\[\affilskip]
$^3$Harvard-Smithsonian Center for Astrophysics, 60 Garden Street, Cambridge, MA 02138, USA}

\pubyear{2013}
\volume{301}  
\pagerange{1--2}
\jname{Precision Asteroseismology}
\editors{J.A. Guzik, W.J. Chaplin, G. Handler \& A. Pigulski, eds.}
\begin{document}

\maketitle

\begin{abstract}

We characterized the time intervals between the interchanges of the alternating high- and low-amplitude extrema of three RV Tauri and three RR Lyrae stars.

\end{abstract}

\firstsection 
              
\section{Introduction}

The brightness variations of RV Tauri stars are characterized by alternating primary and secondary minima. The alternation is not strictly regular; the order of the deep and shallow minima shows occasional reversals \cite[(Wallerstein 2002)]{Wall}. According to the most likely explanation the alternating light variation is caused by the 2:1 resonance between the fundamental and the first-overtone mode \cite[(Fokin 1994)]{Fokin}. In this near-resonance state the system may display low-dimensional chaotic behaviour, as proposed by \cite[Buchler \& Kov\'acs (1987)]{Buchler} and \cite[Buchler et al.~(1996)]{Buchleretal}. The chaos hypothesis is also supported by the observed interchanges of deep and shallow minima.

Similar interchanges were recently discovered in the order of the period-doubled pulsation cycles in RR Lyrae stars observed by the \textit{Kepler} space telescope \cite[(Szab\'o et al. 2010)]{Szabo}. In these cases the period doubling of the fundamental mode is caused by a 9:2 resonance with the ninth overtone. By this time the four-year long photometric data of \textit{Kepler} have become available to study the nature of the interchanges of the alternating maxima in these RR Lyrae stars and to compare to the similar phenomenon in RV Tauri variables.

\section{Data and method}

We analysed three RV Tauri stars (TT Oph, UZ Oph, and U Mon) and three RR Lyrae stars (RR Lyr, V808 Cyg, V355 Lyr) that clearly show the interchanges. RV Tauri light curves are based on several data sources: visual and photometric measurements of amateur astronomers in the AAVSO, AFOEV and VSOLJ databases, data from the ASAS \cite[(Pojma\'nski 1997)]{Poj}, NSVS \cite[(Wo\'zniak et al.~2004)]{Woz} and Catalina Sky surveys \cite[(Drake et al.~2009)]{Drake}, as well as the observations of the \textit{Hipparcos} satellite \cite[(ESA 1997)]{ESA} and the Optical Monitoring Camera aboard the Integral satellite \cite[(Mas-Hesse et al.~2003)]{Mass}. In the case of RR Lyrae stars we used all the available \textit{Kepler} photometric data between the Q1-Q15 runs. 

To determine the extrema of the cycles we applied least-squares parabola and cubic spline fitting. We followed the method used by \cite[Moln\'ar et al.~(2012)]{Molnar} by connecting the even- and odd-numbered cycles respectively. This representation significantly helps to recognize the epoch of the interchanges.

\section{Results and conclusions}

All three RV Tauri stars show interchanges on short and long timescales. The characteristic interval between two interchanges are 14-21 cycles for TT Oph, and 3-9 cycles for U Mon. The longer timescale denote 112-119 cycles for TT Oph and 39-51 for U Mon. Interestingly, the light curve of TT Oph remains in the same phase during the first 249 cycles. U Mon also shows a section of 136 cycles without interchanges. The total number of interchanges is low for UZ Oph, but a short timescale is suspected around 5-25 cycles and a long one around 45-65 cycles.
	
The timescales of the interchanges of RR Lyrae stars are not divided into two groups; instead they spread to a broad interval between 3 and 39, 86 or 60 cycles for V808 Cyg, V355 Lyr and RR Lyr respectively. Most of the intervals are around 6-30 cycles for all RR Lyrae stars. We note that determination of the timescales of the interchanges were not unambiguous for V355 Lyr and RR Lyr, where the alternation seemingly disappears occasionally, making the lengths between the interchanges uncertain up to a few cycles.

No periodicities were recognized in the occurrences of the interchanges for any of the analysed stars. RV Tauri stars can stay for long periods of time in one phase without interchanges. Similar behaviour was not observed in the RR Lyrae stars. The Blazhko periods are $\sim$170 cycles long for V808 Cyg, $\sim$66 cycles for V355 Lyr and $\sim$70 cycles for RR Lyr: the time intervals between the interchanges do not show any correlation with the Blazhko periods.

\section{Acknowledgements}

This work has been supported by the Hungarian OTKA grant K83790, the MB08C 81013 Mobility-grant of the MAG Zrt., the "Lend\"ulet-2009" Young Researchers' Programme of the Hungarian Academy of Sciences and the KTIA URKUT\_10-1-2011-0019 grant. The \textit{Kepler} Team and the \textit{Kepler} Guest Observer Office are recognized for helping to make the mission and these data possible. We acknowledge with thanks the variable star observations from the AAVSO International Database, the AFOEV database, operated at CDS, France, and the VSOLJ database contributed by observers worldwide. KK is grateful for the support of her Marie Curie Fellowship (255267- SAS-RRL, FP7).

\end{document}